\documentclass[9pt]{IEEEtran}

\newtheorem{remark}{Remark}

\usepackage{cite}
\usepackage{amsmath,amssymb}
\usepackage{graphicx,subfig}
\usepackage{hyperref}

\hypersetup{
  colorlinks   = true, 
  urlcolor     = blue, 
  linkcolor    = blue, 
  citecolor   = red 
}

\usepackage{docmute}
\usepackage[table]{xcolor}

\newenvironment{abbreviations}{\begin{list}{}{}}{\end{list}}

\usepackage{tabularx}
\newcolumntype{L}[1]{>{\raggedright\let\newline\\\arraybackslash\hspace{0pt}}m{#1}}
\newcolumntype{C}[1]{>{\centering\let\newline\\\arraybackslash\hspace{0pt}}m{#1}}
\newcolumntype{R}[1]{>{\raggedleft\let\newline\\\arraybackslash\hspace{0pt}}m{#1}}

\usepackage{MathDefs}

\usepackage{verbatim}
\usepackage{tikz}

\makeatother

\begin{document}

\title{An Approach to Track Reading Progression Using Eye-Gaze Fixation Points}

\author{
S. Bottos and 
B. Balasingam, {\em Senior Member, IEEE}
\thanks{Submitted to IEEE Transactions on Instrumentation \& Measurement, January, 2019} 
\thanks{
The authors are with the 
Department of Electrical and Computer Engineering, 
University of Windsor, 401 Sunset Avenue,
Windsor, Ontarion, N9G 3P4, Canada.
E-mail:  \{bottos,singam\}@uwindsor.ca,
Contact TP: +1(519) 253-3000 ext. 5431, 
Fax: +1(519) 971-3695
} 
}

%

\maketitle

\begin{abstract}
In this paper, we consider the problem of tracking the eye-gaze of individuals while they engage in reading.
 Particularly, we develop ways to accurately track the line being read by an individual using commercially available eye tracking devices. Such an approach will enable futuristic functionalities such as comprehension evaluation, interest level detection, and user-assisting applications like hands-free navigation and automatic scrolling. Existing commercial eye trackers provide an estimated location of the eye-gaze fixations every few milliseconds. However, this estimated data is found to be very noisy. As such, commercial eye-trackers are unable to accurately track lines while reading. In this paper we propose several statistical models to bridge the commercial gaze tracker outputs and eye-gaze patterns while reading. We then employ hidden Markov models to parametrize these statistical models and to accurately detect the line being read. The proposed approach is shown to yield an improvement of over 20\% in line detection accuracy.
\end{abstract}


\section{Introduction}

It is no secret that data obtained through the tracking of one’s eye-gaze has the potential to reveal valuable information regarding that individual’s cognitive processes. Indeed, following the pioneering research of Louis Javal's from as early as 1879, psychologist Edmund Huey first began tracking the eye movements of test subjects during the act of reading in order to establish a link between ocular behaviour and cognitive processes (circa 1908) \cite{huey1908psychology} – marking the first era of eye movement research in the context of cognitive process and behavioural inference. It was later postulated in 1998 in a lengthy collection of works and studies in the field of eye movements during reading and informational processing \cite{rayner1998eye} that, due to advances in technology and a rapidly growing interest in the field, the advent of the third era had occurred some twenty years prior following the second era, which was characterized by its focus on applied, experimental psychology. At present day, twenty years later, it may be that we have advanced from the third era to the fourth. Affordable and advanced eye-tracking technology such as the Tobii \cite{tobii} and Gazepoint \cite{gazept} replace their invasive, expensive, and laboratory-condition dependant counterparts of old. Additionally, interest in the potential to decode raw eye-gaze data in order to infer the otherwise hidden cognitive state of the beholder has never been higher – as said advances in technology have provided the means to perform complex experiments with unprecedented accuracy, similar to the manner by which the processing power of modern-day computers has ushered in the age of Big Data \cite{robert2014machine} by improving the capabilities of Machine Learning as a whole \cite{ friedman2001elements}. 

While the study of gaze patterns during the act of reading has been prevalent for over a century, algorithms and systems which are capable of tracking a reader's progression through a block of text remain a rarity. Even state-of-the-art hardware is inherently inaccurate, making it difficult to identify the precise focal point of visual attention\cite{hyrskykari2006utilizing}. The demand for such algorithms and systems, which are capable of inferring a user’s particular focal points amongst noisy data, exists in both academia and industry, and have a number of applications. For instance, content recommender algorithms as well as the placement of content itself may be optimized with the use of reliable focal point information collected from users \cite{granka2004eye, xu2008personalized, puolamaki2005combining} by examining content which is confirmed by eye gaze data to have been seen by a user and not acted upon (ie: not clicked), in addition to content which has been acted upon \cite{zhao2016gaze}. Naturally, this ability to better understand the saliency of visual stimuli remains highly relevant in the realms of cognitive science as it would allow for more accurate study and diagnoses of disorders such as depression and anxiety \cite{sanchez2013attentional, armstrong2012eye}, as well as various autism spectrum disorders \cite{neumann2006looking, klin2002visual}. Human emotions and intentions can be discerned through careful examination of their eye-gaze fixations \cite{rosa2015beyond, florea2018future}. Even detecting the presence of certain types of cancer is made simpler by involving eye-tracking in the diagnostic procedures \cite{lin2018contactless}.

\par Further applications of eye tracking technology exist in the field of human-machine systems \cite{poole2006eye}, particularly with respect to assistive technology, including the development of fully-functional robotic prosthetics \cite{mcmullen2014demonstration} and visual input-based implements to allow humans to interact more naturally with multi-media devices such as computers and televisions \cite{zhang2017eye, hansen1995eye}. In the same vein, brand new fields which seek to optimize the relationship between human and machine, such as the field of Augmented Cognition, have developed as the capability to extract information non-invasively via the eyes advances. Notable emerging technologies include Adaptive Training \cite{coyne2009applying}, in which content, presentation format, and pacing of training modules are modified according to mental fatigue --- measured by observing visual scanning strategies. The United States Navy, similarly, is exploring the potential to decrease operator error during unmanned aircraft missions by detecting operator fatigue using eye-gaze metrics \cite{mannaru2016cognitive}. Furthermore, in the interest of improving public safety, researchers are testing the feasibility of ensuring that vehicle operators keep their eyes on the road with the use of gaze monitoring technology \cite{sodhi2002glance, pohl2007driver, yang2017effect}. Clearly, eye-gaze tracking has a place in the modern world. In this paper we will focus on detecting a horizontal region of focus, such as in line-detection during reading.




  \begin{figure*}
  \centering
  \includegraphics[width=6.2in]{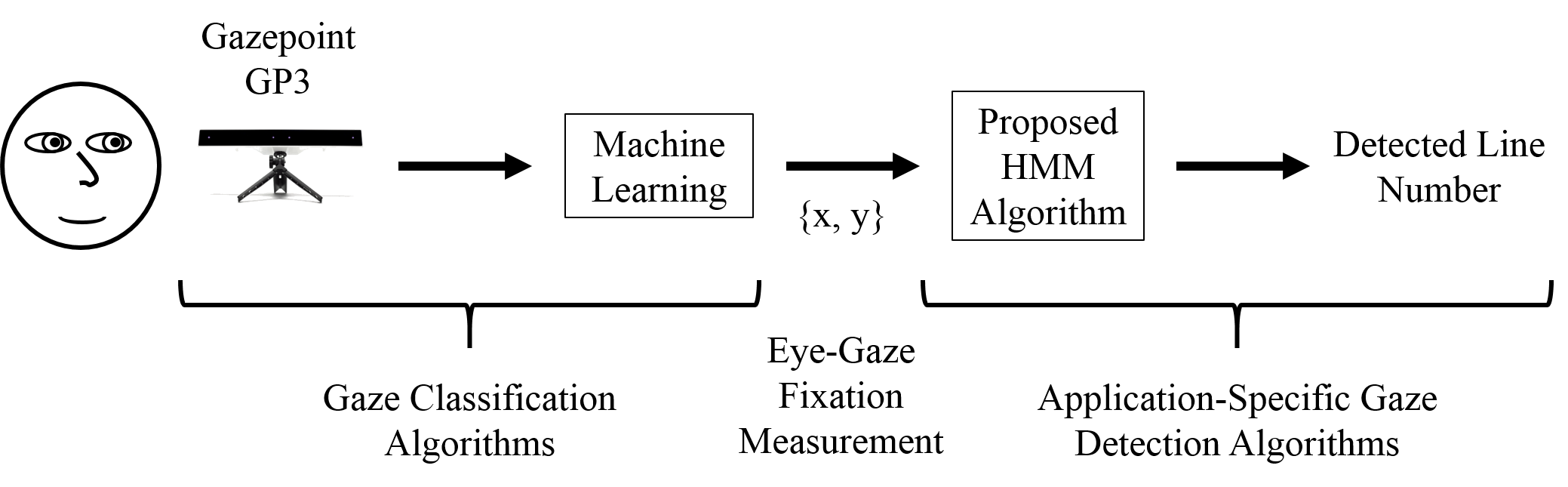}
  \caption{
  {\bf Line detection system (LDS).}
  A block diagram of the proposed LDS. 
  }
  \label{fig:generalalg}
  \end{figure*}
  
It can be seen that eye-tracking technology itself is no new arrival. Fourth-era hardware, such as the Gazepoint GP3 \cite{gazept} used to collect data during the research described in this paper, utilize machine learning processes in order to first recognize the location of a user’s face, then to extract the location of the eyes belonging to the said face. Next, pupillary metrics are measured and interpreted which, after a short calibration step (as is the case with the Gazepoint GP3) in which the dimensions of the user’s field of view are established and translated to a set of x-y coordinates corresponding to the estimated point of gaze within the established parameters. Figure \ref{fig:generalalg} offers a visual representation of this step, referred to as Gaze Classification, along with a second step in which the x-y coordinates are further analyzed using various machine learning techniques ---  an application-specific gaze detection step. Previous efforts have attempted to determine points of interest in films and pictures using clustering algorithms \cite{martinet2008analyzing, naqshbandi2016automatic}, although classic algorithms such as k-means yield mediocre results while more advanced algorithms are computationally expensive. Inaccuracy and computational cost aside, clustering algorithms are not well-suited for real-time, progression-based applications. In this paper, the use of a hidden Markov model-based algorithm in the detection of a user’s progression/regression through distinct horizontal regions of the user’s field of vision is discussed. A useful analogy which will be used throughout this text will be the act of reading. As was previously mentioned, algorithms with such capabilities are rare. Previous research outlines the complex nature of the simple task of reading, such as unpredictable re-visitation of previous passages and the skipping of words\cite{rayner1997understanding, drieghe2005eye}. A similar attempt to track reading progression \cite{paeglis2006maximizing} required the use of sophisticated equipment highly regulated conditions, and filtered data, focusing on the measurement of saccades and lingering gaze points. The hidden Markov model-based algorithm aims to reduce dependency on external conditions, and offer more robust and accurate tracking given noisy data. This is accomplished by first dividing the screen into a configurable number of horizontal regions – which serve as distinct states for the model. In the analogy of reading, each state may represent a single line – or a group of lines – of text. A training sequence must next take place in order to allow the model to learn its optimum parameters given the present environment – in this case, the probabilities corresponding to certain types of behaviour during reading. The model is then able to generalize on previously unseen data in real time in order to accurately track a reader’s progression in the presence of poor quality, noisy data.

\par The remainder of this paper is organized as follows: In Section \ref{sec:problem_def}, the problem to be solved is discussed. Section \ref{PA} describes our proposd approach to line detection while reading. Performance evaluation of the proposed approach was presented in Section \ref{sec:CA} through simulated data. In Section \ref{sec:results_real}, the proposed line detection algorithm is tested using real-world eye-gaze data collected using a commercial eye-gaze tracking device, and the paper is concluded in Section \ref{sec:conclusions}.

\def\midas{In reference to the Greek mythological King who got his wish, of turning everything he touched into gold, granted. }

\section{Problem Definition} \label{sec:problem_def}
The main goal is to detect, using artificial means, a reader's progression through a block of text. With no prior knowledge regarding the layout of the page displayed on screen in terms of text position, text size, and distance between individual lines of text --- the problem to solve becomes one heavily reliant on accurate pattern recognition. If one is to consider a page in a novel, containing multiple lines of unbroken text of equal length and equal spacing, then the act of reading, speaking in terms of non-Semitic languages, will in the ideal case begin at the top-left of the page and track left to right. When one line of text is completed, the reader begins the same left-to-right progression on a new line immediately below that which was previously read. A less ideal assumption, albeit a more reasonable one, would be to consider that the reader may also decide to occasionally return to previously read text rather than progressing in one direction only, or skip forward through text by multiple lines, which must also be kept in mind. Gaze tracking hardware is able to accurately detect the point at which an individual is looking on the screen for which it has been calibrated. By using the output from such a device, it is possible to fashion a "Reading Pattern Detection System" (RPDS).
A ``line detection system (LDS)'' is an integral part of the RPDS, and the focus of this paper. 
\par

The need for a special algorithm for the LDS arises due to two factors: 
\begin{itemize}
\item The {\em measurement noise} introduced  as a result of classification errors made by the gaze tracker's computer algorithms \cite{mannaru2017performance, ramdane2008adaptive}.
\item The {\em MIDAS\footnote{\midas} touch problem} in eye-gaze tracking. It is an intrinsic characteristic of the human eye that it rarely stays focused in one point regardless of where the mind is concentrated \cite{istance2008snap,velichkovsky1997towards}.
\end{itemize}
As a result of the above two sources of noise, the data obtained from the Gazepoint GP3 \cite{gazept} and similar state of the art eye-tracking devices are corrupted by some level of noise. 
In this paper, we consider the above two noise sources together and refer to them as ``gaze measurement noise''. 
The gaze measurement noise is defined by its standard deviation $\sigma_x$ and $\sigma_y$, in the $x$ and $y$ directions, respectively. 
  \begin{figure*}[t]
  \begin{center}
  \subfloat[][$\sigma_x = \sigma_y = 0.02$ line-width]{\includegraphics[width=.4\textwidth]{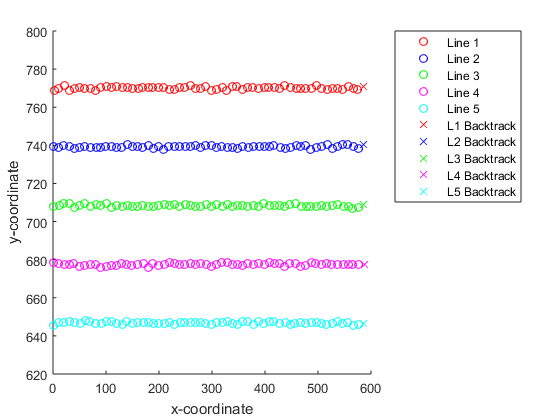} \label{fig:a} } \hspace{20pt}
  \subfloat[][$\sigma_x = \sigma_y = 0.2$ line-width]{\includegraphics[width=.4\textwidth]{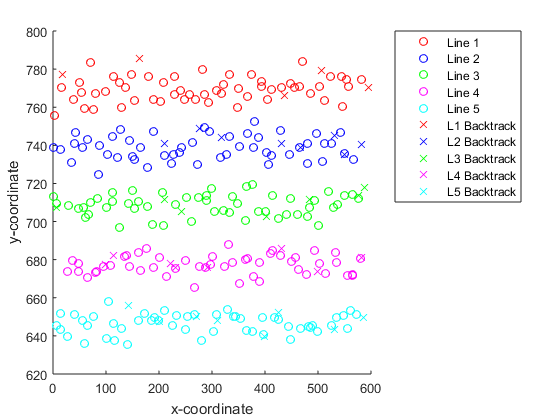} \label{(b)} }\\
  \subfloat[][$\sigma_x = \sigma_y = 0.3$ line-width]{\includegraphics[width=.4\textwidth]{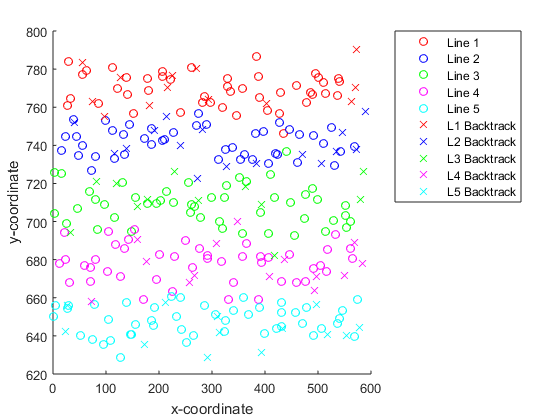} \label{(c)} } \hspace{20pt}
  \subfloat[][$\sigma_x = \sigma_y = 1$ line-width]{\includegraphics[width=.4\textwidth]{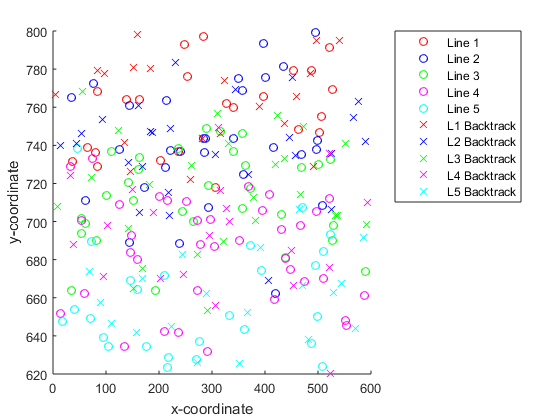} \label{(d)} }\\
  \caption{{\bf Simulated eye-gaze measurements for different measurement noise levels.}
  The $(x,y)$ Coordinates of the eye-gaze point data while reading lines 1 to 5 are shown for different noise levels. 
  The unit of the measurement noise standard deviation is shown in ``line-widths,'' i.e., the distance between two adjacent lines. 
  }
  \label{fig:sim_gp_reading}
  \end{center}
  \end{figure*}  
 Figure \ref{fig:sim_gp_reading} demonstrates simulated eye-gaze measurements at various noise levels, 
the key point to note is that gaze points belonging to a single line tend to be separable to a degree which depends on the level of noise --- with gaze points belonging to a single line tending to overlap as the noise levels, $\sigma_x$ and $\sigma_y$ increase (for the simulated data presented in Figure \ref{fig:sim_gp_reading} and discussed in this paper, $\sigma_x=\sigma_y$). As the amount of overlap between individual lines increases with the increase of noise, the accurate detection of discrete lines of text becomes more challenging. The line-detection algorithm must be robust enough to accurately predict a reader's progression even in the case of input data which is highly corrupted by noise. For this paper, we present ``line-detectability analysis'' for varying measurement noise levels of the eye-tracking device.\par
 
Yet another point to consider is that of the computational expense of the LDS. It is necessary for the system to perform calculations at each time step, as data points are recieved in real time from the gaze tracking hardware, which depending on the hardware itself may be in milliseconds. Thus, the LDS's decision making processes should be designed with maximum efficiency and minimum complexity in mind. Furthermore, since the system must be tracking the reader's progress throughout the duration of their task, a minimal amount of stored memory required for the LDS to reach a decision is desirable. The proposed hidden Markov model approach is well suited for real-time implementation

\section*{List of Notations I}

\begin{abbreviations}
\item[${\rm fix}(t)$] Eye-gaze fixation at time $t$ 
\item[$x_{\rm fix}(t)$] $x$ - coordinate of the eye-gaze fixation at time $t$ 
\item[$y_{\rm fix}(t)$] $y$ - coordinate of the eye-gaze fixation at time $t$ 
\item[$N_l$] Number of lines in a portion of text of interest
\item[$L_j$] $j^{\rm th}$ line in a page/paragraph/section of interest
\item[$o(t)$] Discretized observation or observed state (i.e, observed line number) 
\item[$o(\kappa)$] $\kappa^{\rm th}$ batch of observations $o(\kappa) = [o(1),o(2), \ldots, o(T)]$
\item[$L(t)$] True line of eye-gaze fixation corresponding to the observation $o(t)$
\item[$\hat L(t)$] Estimate of $L(t)$
\item[$S(t)$] True state (i.e. true line number)
\item[$\hat S(t)$] Estimate of $\hat S(t)$
\item[$\bpi$] Prior probabilities 
\item[$\bA$] State transition matrix
\item[$\bB$] Observation matrix
\item[$\sigma_x$] Measurement error standard deviation of the eye-tracking device (in the x-direction)
\item[$\sigma_y$] Measurement error standard deviation of the eye-tracking device (in the y-direction)
\end{abbreviations}

\section{Proposed Approach}
\label{PA}

\subsection{Eye-Gaze Observations and Discretization}
\label{subsec:Discrete}
\begin{figure}
  \centering
  \subfloat[][Surveillance region (area of the text in focus)]{\includegraphics[width=.75\columnwidth]{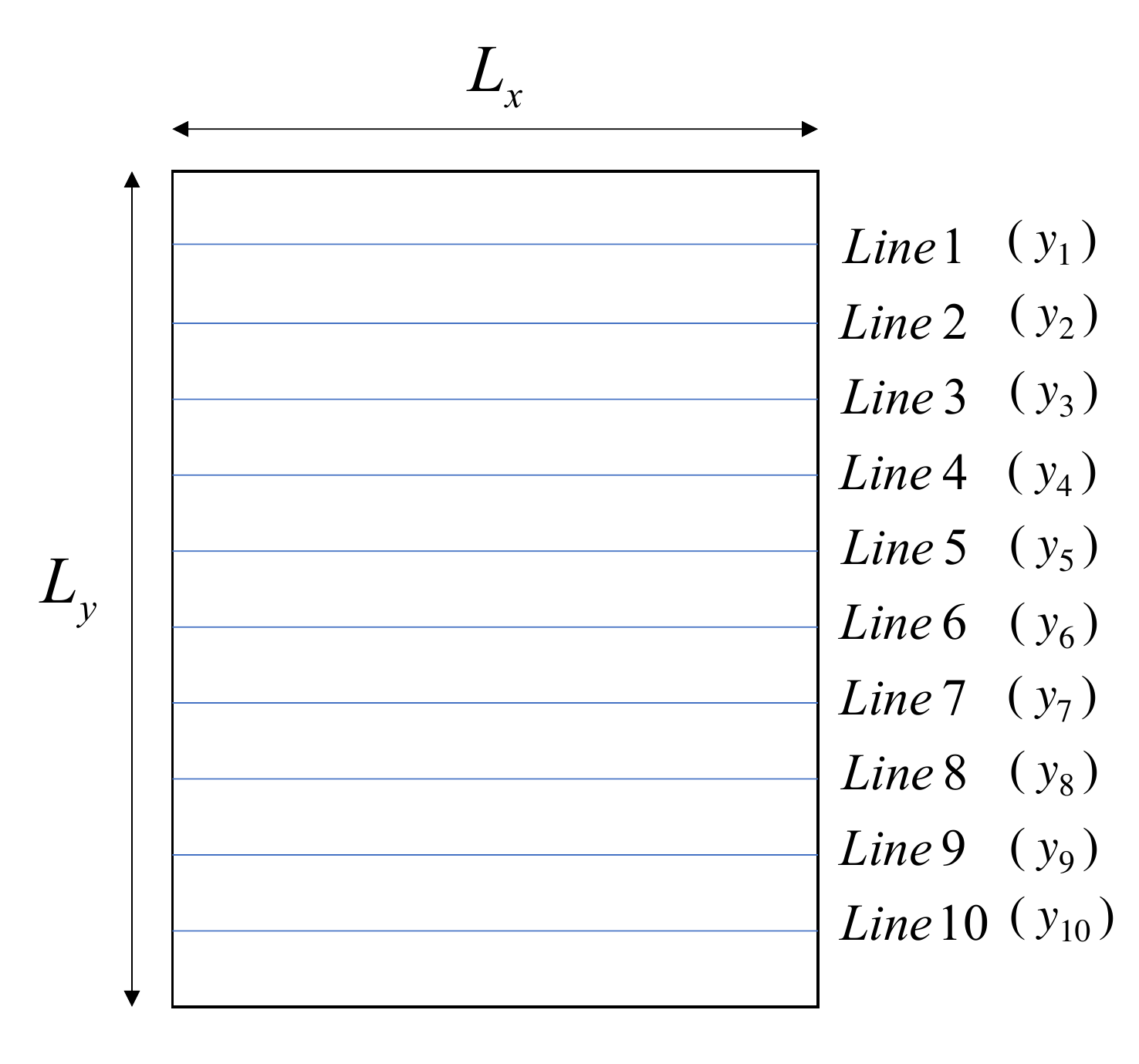} \label{fig:SurvReg}}\\
  \subfloat[][Position of the surveillance region on computer monitor]{\includegraphics[width=.75\columnwidth]{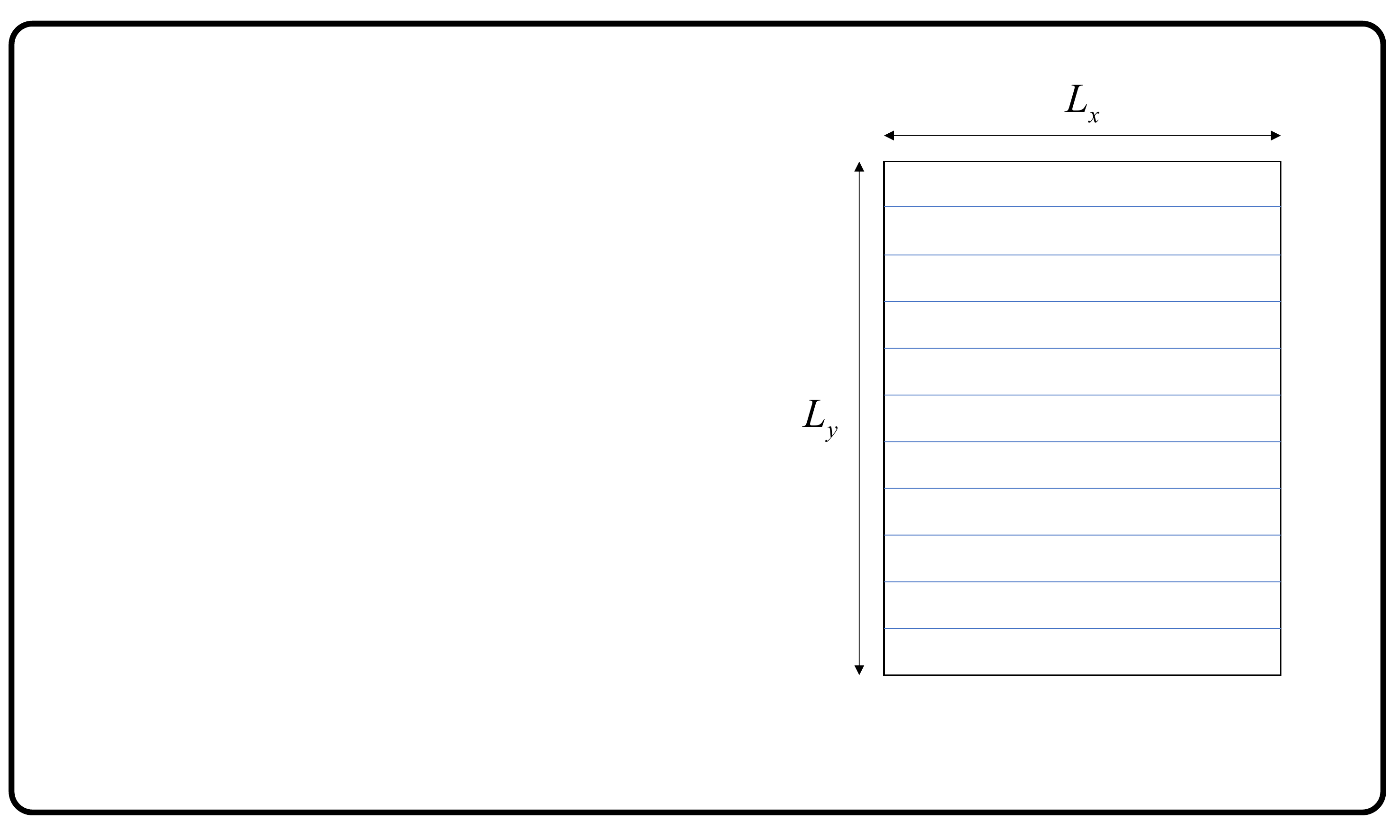} \label{fig:SurvRegB}} 
  \caption{
  {\bf Surveillance region.}
A sample paragraph, with dimensions $L_x \times L_y$, where each line simulates a line of text in an electronic document or web-page. 
The y-coordinate of each line is indicated as $y_1, \ldots, y_{10}.$}
  \label{fig:samplewithdims}
  \end{figure}

The proposed approach in this paper is based on discrete hidden Markov models. 
As such, first we explain how the eye-gaze fixation measurements (obtained from the eye-tracking device) are {\em discretized}, with the help of Figure \ref{fig:samplewithdims}\subref{fig:SurvReg} illustrating a block of text consisting of 10 lines. 
During a reading activity the eye-gaze fixation measurements are produced by an eye-tracking device every few milliseconds; these measurements indicate the point on which the eyes were fixated during that short time interval. 
Let us denote a particular measurement of the eye-gaze fixation obtained at time $t$ as
\begin{eqnarray}
{\rm fix}(t) = [x_{\rm fix}(t), y_{\rm fix}(t)]
\label{eq:GP}
\end{eqnarray}
It is assumed that each measurement ${\rm fix}(t) $ is restricted to the $L_x \times L_y$  surveillance region containing the text, as shown in Figure \ref{fig:samplewithdims}\subref{fig:SurvReg}.
However, in reality, eye-gaze fixation points from outside the surveillance region will be often recorded. 
If the boundaries of the surveillance regions are known, then the measurements that fall outside the surveillance region can be simply discarded, otherwise, the statistical model of the line-detection system can be trained to ignore them. 
In this paper, the surveillance region is always considered to be filled with a block of text, as represented by the lines in Figure \ref{fig:samplewithdims}\subref{fig:SurvReg}.

\begin{figure}[h]
  \centering
  \includegraphics[width=.95\columnwidth]{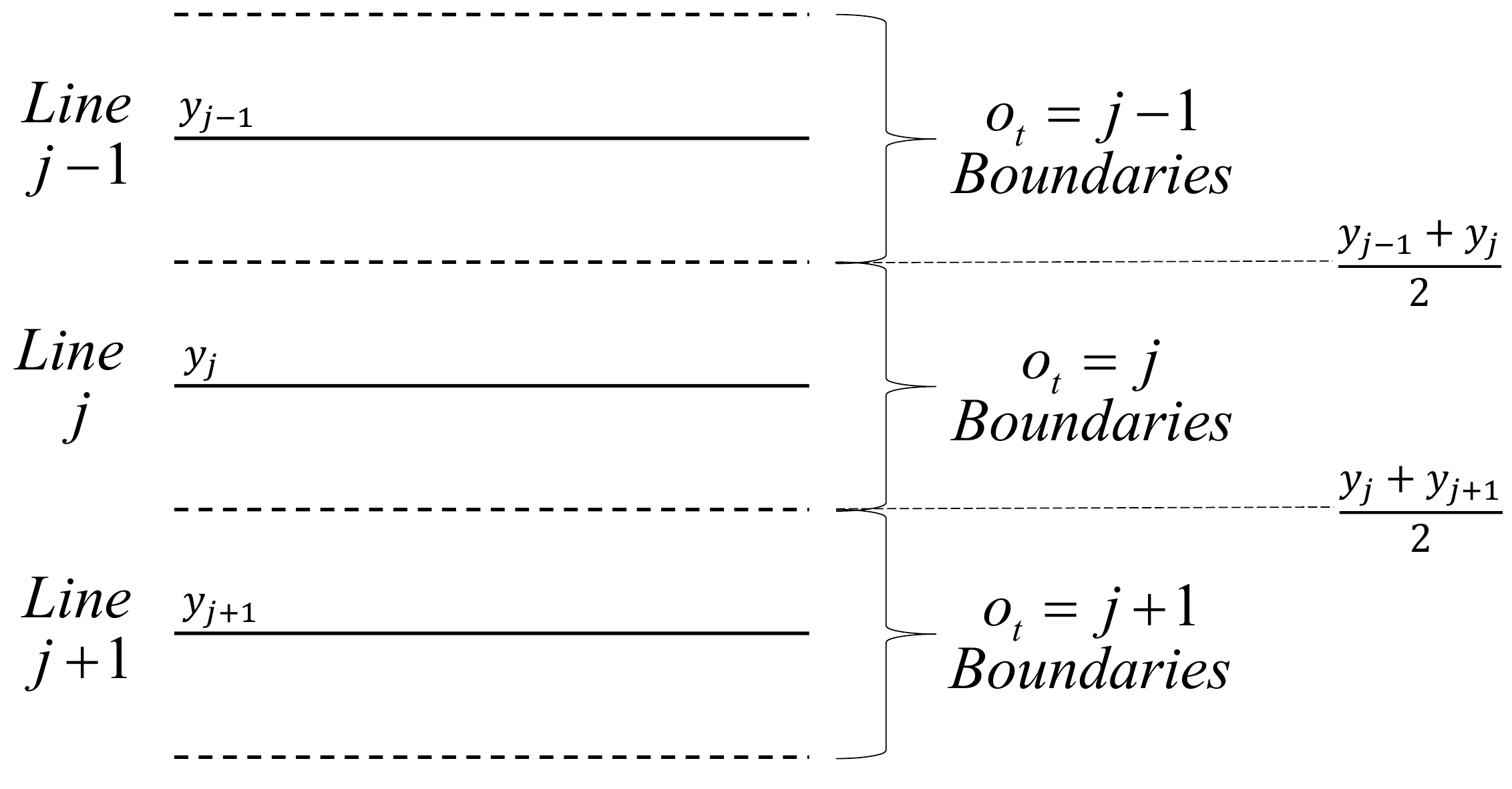}
  \caption{ 
  {\bf Discretization of continuous gaze-point observations.}
  The discretization boundaries are marked using dashed lines. 
  }
  \label{fig:ObsBounds}
  \end{figure}

Let us denote $y_j$ to indicate the the y-coordinate of the $j^{\rm th}$ line as shown in Figure \ref{fig:samplewithdims}\subref{fig:SurvReg}, each of which is determined by dividing the length of the page $L_y$ evenly --- under the assumption that text is distributed evenly throughout the area which it occupies. 
Now, The objective of the discretizer is to classify each gaze measurement ${\rm fix}(t)$ in terms of line number; this can be simply done as follows (also, see Figure \ref{fig:ObsBounds}):
\begin{eqnarray}
o(t)  = \left\{ j  \left|  \frac{y_j + y_{j-1}}{2}   < y_{\rm fix}(t) < \frac{y_j + y_{j+1}}{2}   \right.   \right\} 
\end{eqnarray}
where $o(t)$ is the discretized observation corresponding to the eye-fixation measurement ${\rm fix}(t)$ at time $t$. 
It must be stressed that, in this paper, we only consider the $y$ - observations since our objective is to track the line on which the eye-gaze fixation is focused while reading.

  From the discretization procedure described above, it must be clear that
\begin{eqnarray}
o(t)  \in \{1, \ldots,  N_l \}
\end{eqnarray}
where $N_l$ denotes the number of lines in a page of interest. 
For simplicity, we  will assume that $N_l$ is the same in each page considered for training and testing;
later we will discuss how this assumption can be easily relaxed, in Section \ref{sec:conclusions}. 

Now, the LDS problem can be formally stated as follows:
Given a batch of $T$ discretized observations $ o(t)$, where $t = 1, \ldots, T$ and $o(t)\in \{1, \ldots, N_l \}$, estimate the corresponding lines on which a reader was focused, i.e., obtain the estimates $\hat  L(t)$ for $t = 1, \ldots,  T$ where  $\hat  L(t) \in \{1, \ldots, N_l \}$.


\subsection{Discrete State-Space Model}
\label{subsec:DSSM}

Let us denote the true line on which a reader was focused at time $t$ as $S(t)$ where $S(t) \in \{1, \ldots, N_l \}$.
The true line of focus at time $t+1$, $S(t+1) \in \{1, \ldots, N_l \}$, is dependent on the present state $S(t)$ as well as all previous states $S(t-1), S(t-2), \ldots, S(1) $, i.e.,
\begin{eqnarray}
S(t+1)  = f( S(t), S(t-1), \ldots, S(1), S(0))
\label{eq:process_model}
\end{eqnarray}
Under the Markov assumption, this dependance can be relaxed to the previous state only -- resulting in the following {\em process model}:
\begin{eqnarray}
S(t+1)  = f(S(t))
\label{eq:process_model}
\end{eqnarray}
Considering that each true state takes only discrete values $ \{1, \ldots, N_l \}$, the above process model can be stated as an $N_l \times N_l$  {\em state transition matrix} $\bA$ where the $(i,j)^{\rm th}$ element of $\bA$ can be written as 
\begin{eqnarray}
{a}_{ij} &=& P\left\{ S(t+1) = j|S(t)=i \right\} 
\end{eqnarray}

Each observed data $o(t)$ relates to the true state through a nonlinear function $g(\cdot)$, i.e.,
\begin{eqnarray}
o(t) = g(S(t))
\end{eqnarray}
Now, considering that {\em both} $o(t)$ and $S(t)$ take only discrete values, i.e.,  $o(t)\in \{1, \ldots, N_l \}$ and $S(t) \in \{1, \ldots, N_l \} $, the above observation model can be stated as the following $N_l \times N_l$  {\em observation matrix} $\bB$ where the $(i,j)^{\rm th}$ element of $\bB$ can be written as 
\begin{eqnarray}
{b}_{ij} &=& p\{ S(t)=i| o(t)=j \}
\end{eqnarray}

\begin{figure*}
\centering
\includegraphics[width= .95\textwidth]{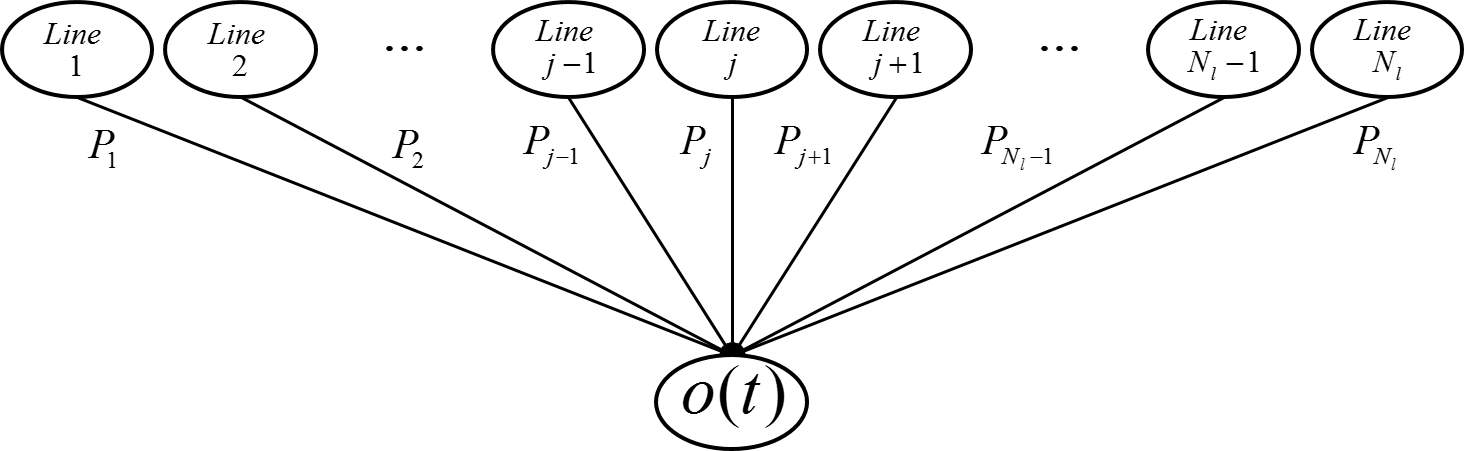}
\caption{
{\bf Observation matrix.}
Given an observation $o(t)$, the observation matrix shows the probability of each possible state $\{1, \ldots, N_l \}$ (i.e., line number) to be the true state. 
}
\label{fig:ObsPath}
\end{figure*}

 It is easy to see that the discrete nature of the observation (observed line number) and state (true line number) allow the problem to be posed as a discrete hidden Markov model (HMM) \cite{rabiner1990tutorial}. 
 In the next sub-section, we describe a training procedure to estimate the parameters of the proposed HMM. 
 Later, these HMM parameters will be used by the proposed Line Detection System (LDS).

\subsection{Proposed LDS Based on Discrete Hidden Markov Models} 

The proposed line detection system (LDS) consists of two important steps:
\begin{enumerate}
\item {\em Parameter estimation.}
The objective here is to estimate the parameters of the discrete state-space model defined in Subsection \ref{subsec:DSSM} to represent eye gaze movement while reading. 
Specifically, the objective is to estimate, $\bpi, \bA,$ and $\bB$, the parameters of the discrete HMM. 
This step is also referred to as {\em training stage} or simply as {\em HMM training.}
\item {\em State estimation.}
Once the (HMM) parameters are estimated, the proposed LDS becomes functional. 
We test its functionality by feeding the discretized observations $o(t)$ with an objective of estimating the states $S(t)$, based on the notations defined in Subsection \ref{subsec:DSSM}.  
In this paper, we also refer to the state estimation step as {\em testing stage} or simply as {\em HMM testing.}
\end{enumerate}

\begin{figure}[h]
  \centering
  \includegraphics[width=.4\textwidth]{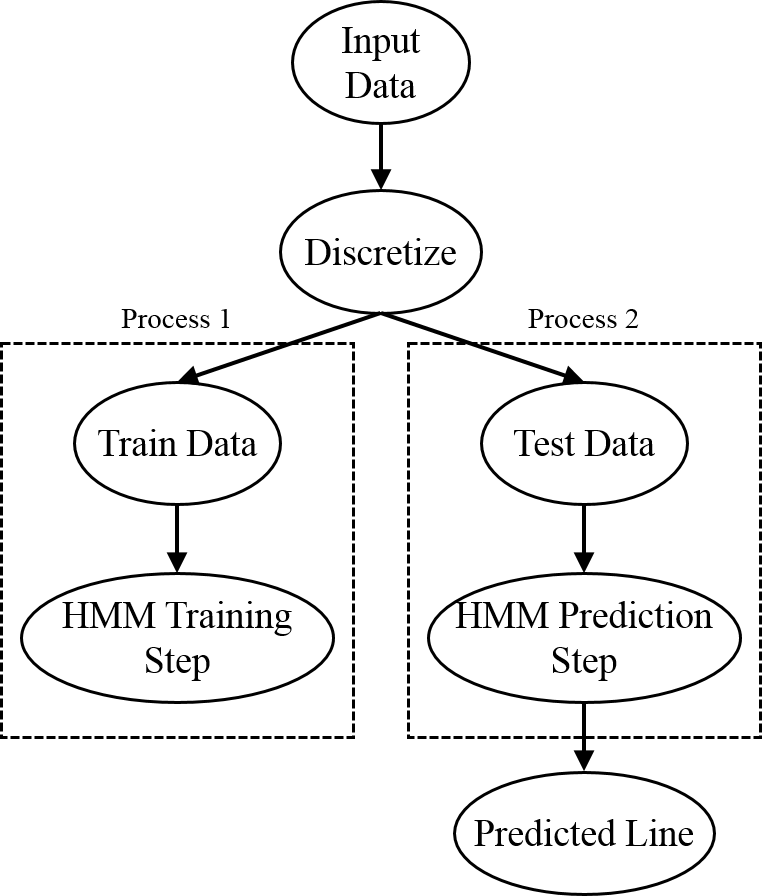}
  \caption{
  {\bf Data collection and testing process.}
  A graphical representation of the individual processes, and flow, of the RPDS. Process 1 must be performed prior to Process 2 in order to fit model parameters to incoming data.
In order to ensure robust performance, the data for the training stage (Process 1) is custom simulated; details can be found in Section \ref{sec:CA}. 
  }
  \label{fig:process}
  \end{figure}

 Figure \ref{fig:process} describes the data flow during the parameter estimation (training) and state estimation (testing) processes. 
 We describe these two processes in detail in the next two subsections. 
 
  
\subsubsection{Parameter estimation}

The objective of the HMM training module is to estimate the model parameters $\bpi, \bA$ and $\bB.$
For training it is assumed in this paper that the dimension of the surveillance region (see Figure \ref{fig:samplewithdims}) and the exact $y$ - coordinates of each lines are known or estimated. It is also assumed that these two quantities (i.e., the dimension of the surveillance region and the exact location of the y-coordinates of the lines) did not change during training, nor did they later during testing. 
This seemingly strict assumption must be adjusted in a practical application --- see Subection \ref{GeneralizedApp} for a discussion on this.

Discrete HMM training is a well studied subject, and the Baum-Welch algorithm \cite{rabiner1990tutorial} provides the best known approach to train an HMM with the use of the expectation maximization (EM) technique.  
Several software platforms, including Matlab, provide optimized HMM training libraries. 
The training process requires an initial guess for the parameters, after which it updates them recursively. 
Assuming the initial parameters to be $\bpi_0,  \bA_0,$ and  $\bB_0$, the trained HMM parameters 
$\bpi, \bA,$ and $\bB$ are obtained as follows:
  \begin{eqnarray}
  [\hat \bpi, \hat \bA, \hat \bB] = {\tt hmmtrain}( \bpi_0,  \bA_0, \bB_0) 
  \end{eqnarray}
where the hat-notation, $e.g., \hat \bpi,$ indicates that the parameter is estimated, and ${\tt hmmtrain}$ represents the Baum Welch algorithm \cite{rabiner1990tutorial} in Matlab. Once convergence is achieved one can set $\bpi = \hat \bpi$, $\bA = \hat \bA$, and $\bB =  \hat \bB$.
 Several other leading machine learning platforms also provide optimized routines for Baum-Welch learning, e.g., Python's scikit-learn package.

A good initial guess is often conducive to fast and accurate convergence during HMM training. 
Next, we describe an approach to select suitable initial parameters  $\bpi_0,  \bA_0,$ and  $\bB_0$ for the proposed HMM training step. 
The selections described below are intended for $N_l = 10$ lines; this can be extended to any number of lines.   

First, the prior probabilities can be selected by taking into account the fact that, most of the time, the reader will begin with the first line of text:

\begin{footnotesize}
\begin{eqnarray}
\bpi_0^T = 
\left[
\begin{array}{cccccccccc}
.91 & .01 & .01	& .01 & .01 & .01 & .01 & .01 & .01 & .01 
\end{array}
\right]
\end{eqnarray}
\end{footnotesize}

The transition probabilities are selected based on the intuitive observation that the transition between two adjacent gaze fixation points ${\rm fix}(t)$ and ${\rm fix}(t+1)$, mostly occur within the same line. The next most common transitions will intuitively occur between the immediate-next or immediate-previous lines;
immediate-next line transitions occur when the person progresses from one line to another in natural progression; and the 
immediate-previous line transition occurs due to backtracked reading of lines. 
Assuming that the probability of immediate-next line transition and the immediate-previous line transition are the same, and the transition probability is zero for gaze-transitions separated by more than one line, the initial guess of the transition probability is given as follows:

\begin{footnotesize}
\begin{eqnarray}
\bA_0 = 
\left[
\begin{array}{cccccccccc}
.9 & .1 & 0	& 0 	& 0 & 0 & 0 & 0 & 0 & 0\\
.05 & .9 & .05  	& 0  	& 0 & 0 & 0 & 0 & 0 & 0\\
0 & .05  & .9 & .05    &  0 & 0 & 0 & 0 & 0 & 0\\
0 & 0 & .05  & .9 & .05    &  0 & 0 & 0 & 0 & 0\\
0 & 0 & 0 & .05  & .9 & .05    &  0 & 0 & 0 & 0\\
0 & 0 & 0 & 0 & .05  & .9 & .05    &  0 & 0 & 0\\
0 & 0 & 0 & 0 & 0 & .05  & .9 & .05    &  0 & 0\\
0 & 0 & 0 & 0 & 0 & 0 & .05  & .9 & .05    &  0\\
0 & 0 & 0 & 0 & 0 & 0 & 0 & .05  & .9 & .05  \\
0 & 0 & 0 & 0 & 0 & 0 & 0 & 0 & .1 & .9  \\
\end{array}
\right]
\end{eqnarray}
\end{footnotesize}

\def\signote{We will discuss in terms of $\sigma_y$ for brevity and since the main concern of the LDS is the $y$ - measurement of each gaze fixation point, but keep in mind that for simulated data $\sigma_y$ = $\sigma_x$} 

The true gaze corresponding to a certain fixation point observation ${\rm fix}(t)$ (more specifically, the discretized fixation point observation $o(t)$) can be anywhere on the page depending on the {\em measurement error standard deviation} of the eye-tracking device.
For example, Figure \ref{fig:sim_gp_reading}\subref{fig:a} indicates a scenario where the measurement error standard deviation\footnote{\signote} of the eye-tracking device is very low; and Figure \ref{fig:sim_gp_reading}\subref{(d)} indicates a scenario where $\sigma_y$ is very high. 
When $\sigma_y$ is very low the observation matrix $\bB$ approaches an identity matrix i.e., it is (almost) very likely that $o(t)$ corresponds to the true line number $S(t) = o(t)$.
On the other hand, when  $\sigma_y$ is very high, the observation matrix $\bB$ takes a Toeplitz form. 
Figure \ref{fig:ObsPath} demonstrates this further: with each line \textit{j} having the ability to generate \textit{o(t)} with some probability \textit{P\textsubscript{j}}, intuitively, one can say that the greatest probability occurs at \textit{P\textsubscript{j}} when \textit{o(t)} =  \textit{j}, ie:
  \begin{eqnarray}
  {P_j} > {P_{j+1}} > {P_{j+2}} > ... >  {P_{N_l}} \text{ ... when $o(t) = j$}
  \end{eqnarray}
and,
  \begin{eqnarray}
  {P_j} > {P_{j-1}} > {P_{j-2}} > ... >  {P_{1}} \text{ ... when $o(t) = j$}
  \end{eqnarray}
Below, we illustrate an initial guess for $\bB_0$ that is a simplified version derived from the above discussion where we assume $P_1=P_2=...=P_{j-1}=P_j=P_{j+1}=P_{N_l}$

\begin{small}
\begin{eqnarray}
\bB_0 = 
\left[
\begin{array}{cccccccccc}
.91 & .01 & .01	& .01 	& .01 & .01 & .01 & .01 & .01 & .01\\
.01 & .91 & .01 	& .01  	& .01 & .01 & .01 & .01 & .01 & .01\\
.01 & .01 & .91 & .01   &  .01 & .01 & .01 & .01 & .01 & .01\\
.01 & .01 & .01 & .91 & .01   &  .01 & .01 & .01 & .01 & .01\\
.01 & .01 & .01 & .01 & .91 & .01   &  .01 & .01 & .01 & .01\\
.01 & .01 & .01 & .01 & .01 & .91 & .01   &  .01 & .01 & .01\\
.01 & .01 & .01 & .01 & .01 & .01 & .91 & .01   &  .01 & .01\\
.01 & .01 & .01 & .01 & .01 & .01 & .01 & .91 & .01   &  .01\\
.01 & .01 & .01 & .01 & .01 & .01 & .01 & .01 & .91 & .01 \\
.01 & .01 & .01 & .01 & .01 & .01 & .01 & .01 & .01 & .91  \\
\end{array}
\right]
\end{eqnarray}
\end{small}

\noindent where it is assumed that each the off diagonal elements are the same. It must be noted that this is a special case of the Toeplitz matrix.


\subsubsection{State Estimation}

Once training is complete and the parameters $\bpi,\bA,\bB$ are obtained, a batch of observed data sequences 
\begin{eqnarray}
\bo(\kappa) =  \left[ o(1), o(2), \ldots, o(T) \right] 
\end{eqnarray}
can be used to estimate the lines on which {the reader's} eye-gaze was fixated corresponding to each observation $o(t)$. 
Since a batch of $T$ observations is considered, this proposed approach falls under the batch estimation category
\begin{eqnarray}
  \hat \bS(\kappa) = {\tt decode}(\bpi,\bA,\bB,\bo(\kappa)) 
  \end{eqnarray}
where the `decode' routine is carried out based on the Viterbi algorithm \cite{rabiner1990tutorial}. 

\begin{remark}[Computational complexity]
The complexity of the Viterbi algorithm is $\cO(N^2 T)$ where $N$ is the number of states and $T$ is the length of the sequence. 
Assuming that the approximate reading speed is 2 seconds/line and that the sampling rate of the eye-tracking device is $60$ Hz, waiting for 1200 samples, i.e. $T=1200$, will have gaze-data spanning approximately 10 lines. The complexity of decoding will be in the order of 12k flops. 
\end{remark}

\begin{remark}[Constraints for training and testing]
In order to select an arbitrary length of data $T$, it must be assumed that the $y$ -coordinate of each line remains the same as what it was during training. 
This might be a rather unrealistic assumption in {\em free-style reading}, such as, reading PDF documents, Word documents and websites in a computer. 
However, the proposed assumption is valid in custom applications such GUIs and e-books on tablets. Further, the proposed approach can be easily modified to make it applicable in free-style reading. Some details regarding such modifications are discussed in Subection \ref{GeneralizedApp}. 
\end{remark}

\subsection{Generalized Application of LDS}
\label{GeneralizedApp}
\begin{figure}
  \centering
  \subfloat[][The area of interest, shown with stray gaze points which would skew the results of the discretizer]{\includegraphics[width=.45\textwidth]{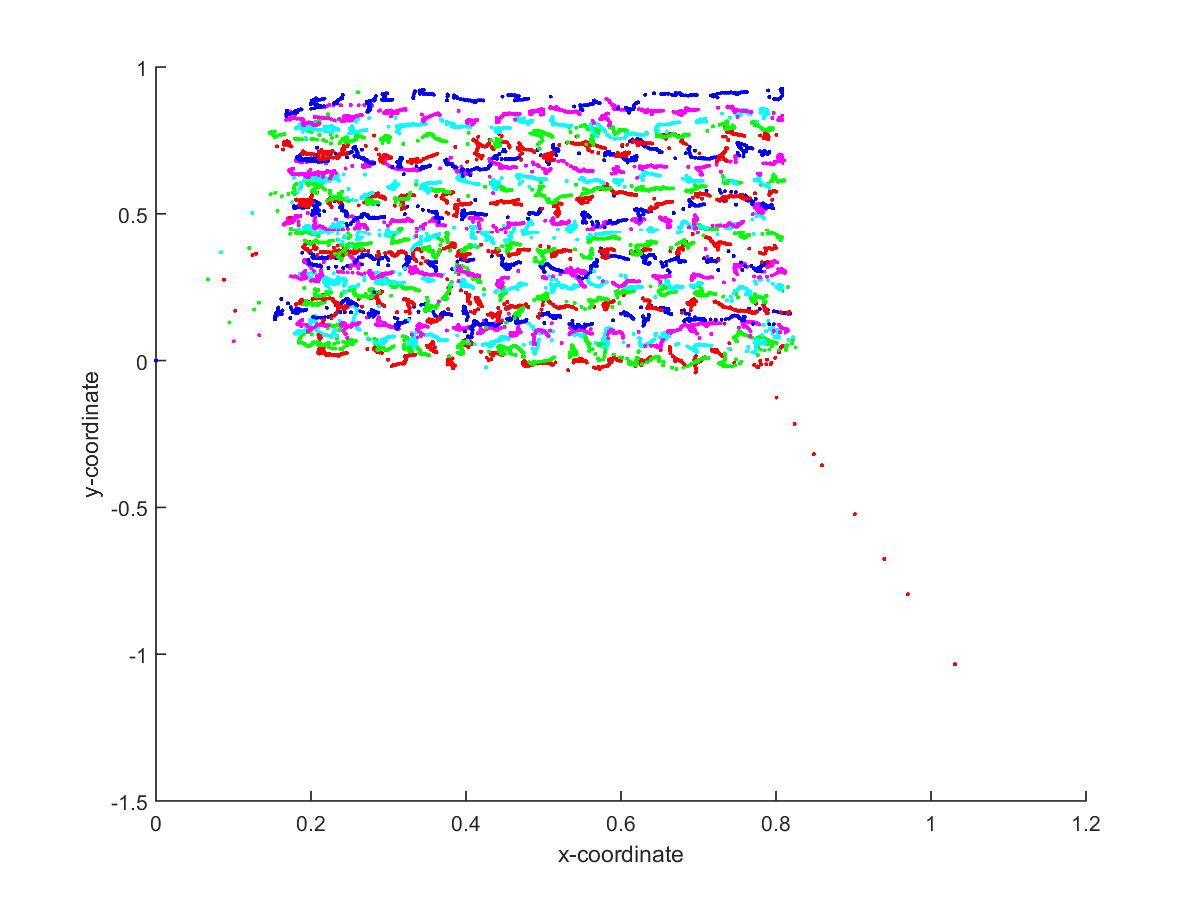}
  \label{fig:uncleaned}}\\
  \subfloat[][The surveillance area after the removal of outliers. Relatively accurate discretization is now possible.]{\includegraphics[width=.45\textwidth]{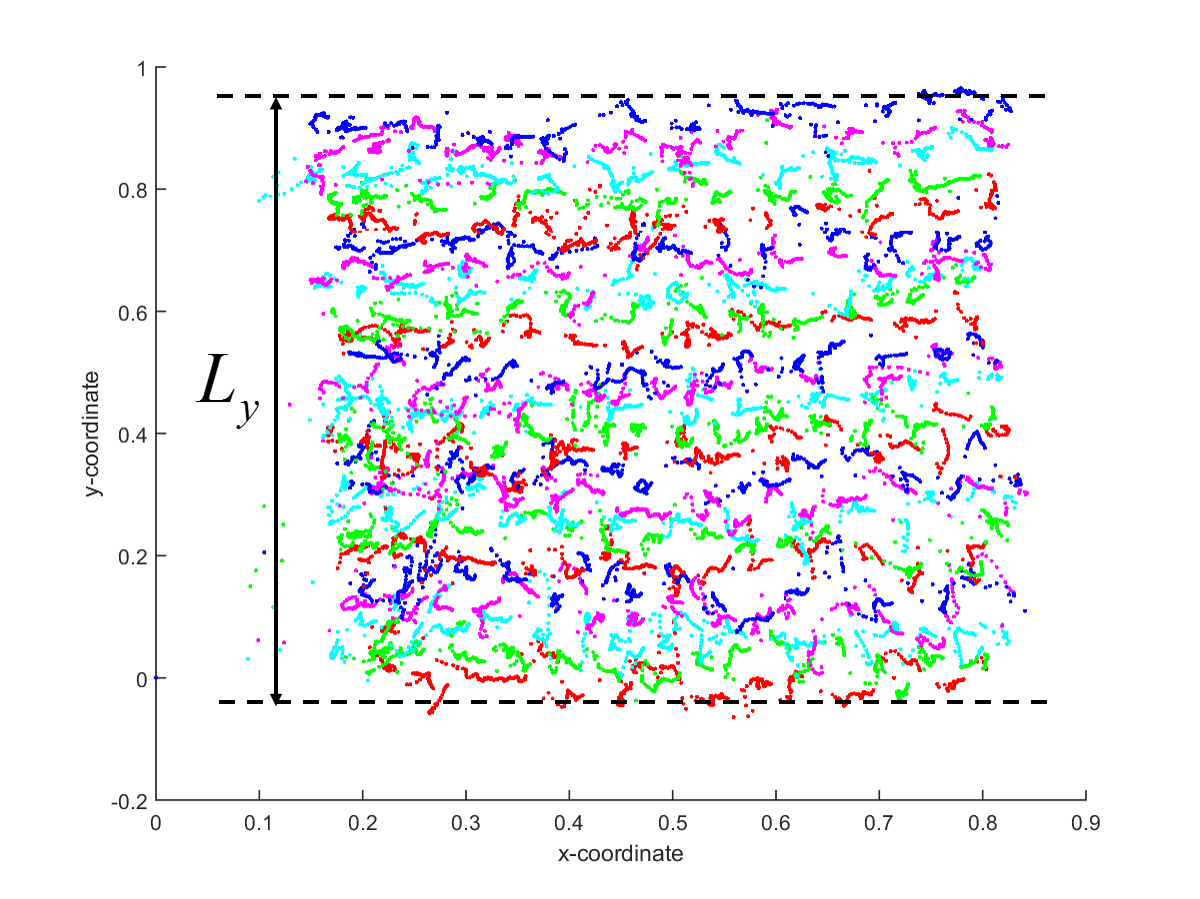} 
  \label{fig:cleaned}} 
  \caption{
  {\bf Surveillance region detection.} After the discarding of outliers, a batch of both maximum and minumum points may be collected in order to establish the range of $L_y$, after which discretization can be performed.
  }
  \label{fig:surveillance_area_estimation}
  \end{figure}

In this sub-section, we discuss an approach of estimating the surveillance region based on collected eye-gaze fixation points only, without knowing explicitly where on the screen the surveillance area is located. It was previously discussed in Section \ref{subsec:Discrete} that, given a known dimension $L_y$, the location of each line on the page $y_j$ could be determined. Let us refer back to Figure \ref{fig:samplewithdims}\subref{fig:SurvRegB}. Imagine a scenario such as this, where data are collected not only within the parameters of the surveillance area but are also scattered across the region of the computer monitor. Indeed it was observed that, in the case of the real data, gaze fixation points were collected not only outside of the surveillance region but outside of the region for which the eye tracker was calibrated --- if an individual's gaze falls anywhere within the detectable field of the tracker, it will register as a data point. As such, it is necessary to discard outliers and establish the true region of interest. Recall that the current model detects only vertical progression, as such $L_y$ only is required for our purposes.

  The assumption was made that, during reading, the vast majority of fixation points will fall in the area of the text which is being read. With this intuition, a cleaning step was performed for each collected page of real data. This cleaning step involved first discarding all points which fell outside of 1.9 standard deviations from the average of all points, followed by averaging a batch of maximum and minimum gaze points in order to establish the upper and lower bounds of $L_y$. Once an estimate for $L_y$ was obtained, discretization could be performed as usual to estimate the location of each line of text, $y_j$. A demonstration of the effect of cleaning is illustrated in Figure \ref{fig:surveillance_area_estimation}.

\section{Computer Analysis}
\label{sec:CA}
In this section, we present a performance analysis of the proposed Line Detection System (LDS). First, we present an objective performance analysis using a simulated eye-gaze data obtained through a simulation that is designed to mimic typical reading; the details are summarized in the remainder of this section. 
Then, we present the performance analysis of the LDS on eye-gaze data collected using a Gazepoint eye-tracker \cite{gazept} collected while a human-subject was reading several pages of text; the details of this analysis are presented in Section \ref{sec:results_real}.

\subsection{Simulated Input Data}
\label{sec:results_simulated}
A set amount of simulated data was first generated, designed to mimic a block of text in the format illustrated in Figure \ref{fig:samplewithdims}.  We will first introduce some key notations which have not been previously defined.

\subsubsection*{Simulation Parameters}

\begin{abbreviations}
  \item [$T_l$] Amount of time a reader will spend on one line of text
  \item [$T_r$] Amount of time a reader will take to return to the beginning of a new line after completing the previous line
  \item [$f_s$] Sampling frequency
  \item [$T_s$] Sampling time, i.e., $T_s = {1}/{f_s}$
  \item [$N_{\rm pages}$] The number of pages for which to generate data
  \item [$N_p$] The number of eye-gaze fixations per page to generate, computed as
  \begin{eqnarray}
  N_p &=& \left(\frac{T_l}{T_s}\right)N_l
  \end{eqnarray}
  \item [$\sigma$] Measurement noise standard deviation on both directions, i.e., $\sigma_x = \sigma_y = \sigma$
  \item [$\bf{o}\it_{p}$] The observation sequence generated for page $p$, with each individual observation in the sequence denoted as $o_p(t)$, for $p \in  \left\{1, ..., N_{\rm pages}\right\}$ and $t =  \left\{1, ..., N_p\right\}$
\end{abbreviations}

\begin{remark}
The eye-gaze fixations were generated by assuming that the progression of reading along each line of text happen at the same, constant rate for every line. 
Even though this is an idealistic assumption, it serves our purpose: to obtain benchmark of performance metrics in order to compare  with the LDS performance in real data. 
\end{remark}

Simulated data was generated by assigning the following values to the previously defined variables (recall that $N_l$ was defined earlier as the number of lines per page),
\begin{itemize}
  \item $T_l$ = 1 ${\rm sec}$
  \item $T_r$ = 0.1  ${\rm sec}$
  \item $f_s$ = 60  ${\rm Hz}$
  \item $N_{\rm pages}$ = 50
  \item $N_{l}$ = 25
  \item $N_p = \left(\frac{T_l}{T_s} \right)N_l = 1\times60\times25=1500$
  \item $\sigma = \left\{1, 0.63, 0.46, 0.37, 0.3, 0.26, 0.25, 0.22, 0.2\right\}$ 
\end{itemize}
where the unit of $\sigma$ is in line-widths. 

\begin{figure}
  \centering
  \includegraphics[width=.4\textwidth]{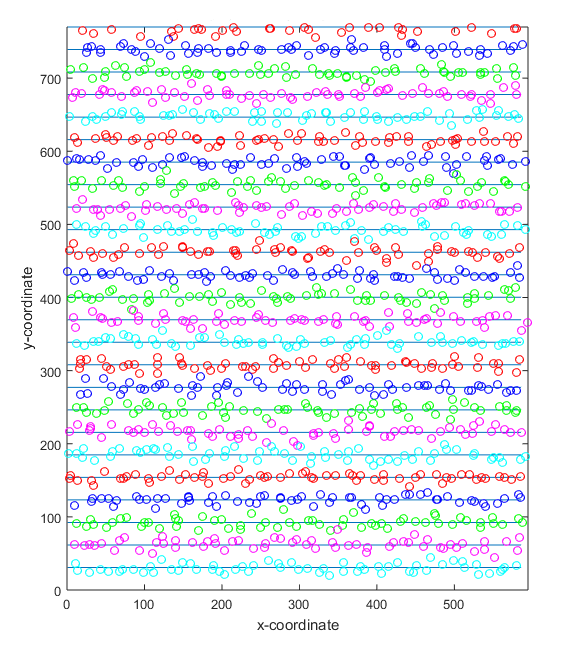}
  \caption{{\bf Full page data collected at $\sigma$ = 0.2.} A sample page of simulated data generated according to the parameters discussed.}
  \label{fig:simdata}
  \end{figure}

  Data in the format defined in equation \ref{eq:GP} is fed to the descretize function as a column matrix, where the column vectors containing all elements $x_{\rm fix}(t)$ and $y_{\rm fix}(t)$  will be referred to as \textbf{x} and \textbf{y} respectively. For visualization purposes, a full sample page of data points generated using $\sigma = 0.2$ is shown in Figure \ref {fig:simdata}. 
  
%
%
%

\par Using the fifty pages of data generated at each noise level, pages 1-40 were reserved for training while the final 41-50 pages were set aside for testing. In other words, the test observation sequences and their resultant predicted state sequences at each noise level were defined as $\bf{o}\it_{test,p}$  and $\bf{s}\it_{p}$, respectively, for $p=$ \{41:50\}. The simulated data was designed to contain labels --- which were of course stripped from the data prior to training and testing --- which indicated the specific line number that a data point had been generated from. Thus, the ground truth values of each true hidden state at each time step were known, and it was possible to quantify the error associated with a predicted state sequence through comparison, which was computed as,
\begin{eqnarray}
e_p &=&  \left[\frac{\sum_{t=1}^{t = N_p}\hat S_p(t) \neq S_p(t)}{N_p}\right]\times 100
\end{eqnarray}
In other words, $e_p$ is the percentage of incorrect line predictions stored in $\bf{s}\it_{p}$. The average error across all test sets, $e_{avg}$ was then computed as,
  \begin{eqnarray}
e_{avg} &=& \frac{\sum_{p = 41}^{50}e_p}{10}
  \end{eqnarray}
In such a manner, $e_{avg}$ was obtained for each noise level, and was repeated for two scenarios --- Non-Randomly Repeated Lines, and Randomly Repeated Lines --- which are explained in their subsequent sections.

\subsection{Reading with No Repetition of Lines} 
In this case, each line of text was generated exactly once, which was intended to simulate a human reader progressing uni-directionally through a block of text with no intentional backtracking, or re-reading of a certain line of text, besides the inherent noise produced with the data. Table \ref{table:noranderrors} presents the average error resulting from ten sets of test data. As expected, as the level of noise increases, the number of incorrect line predictions increases.
\begin{table}[h]
\caption{Average Line Detection Errors, Non-Randomly Repeated Lines}
\label{table:noranderrors}
\begin{center}
\begin{tabular}{|c|c|}
\hline
\rule{0pt}{12pt}\bf{Noise Level}, $\sigma$ & \bf{Error}, $e_{avg}$ \\
\hline
\rule{0pt}{12pt}1 & 66.45 \%\\
\hline
\rule{0pt}{12pt}0.63 & 15.46 \% \\
\hline
\rule{0pt}{12pt}0.46 & 2.48 \% \\
\hline
\rule{0pt}{12pt}0.37 & 0.97 \% \\
\hline
\rule{0pt}{12pt}0.3 & 0.71 \% \\
\hline
\rule{0pt}{12pt}0.26 & 0.30 \%\\
\hline
\rule{0pt}{12pt}0.25 & 0.18 \% \\
\hline
\rule{0pt}{12pt}0.22 & 0.04 \%\\
\hline
\rule{0pt}{12pt}0.2 & 0.05 \% \\
\hline
\end{tabular}
\end{center}
\label{Err}
\end{table}

\subsection{Reading with Random Repetition of Lines} 
In the case of randomly repeated lines, each line of text was generated consecutively a random number of times, ranging from one to five. This simulation was designed to mimic the act of a reader intentionally backtracking or re-reading a specific line a random number of times before proceding to the next line of text, introducing an element of unpredictability to the data. Table \ref{table:randerrors} presents the average error resulting from ten sets of test data containing randomly repeated lines. Once again, as expected, as the level of noise increases, the number of incorrect line predictions also increases.
\begin{table}[h]
\caption{Average Line Detection Errors for Randomly Repeated Lines.}
\label{table:randerrors}
\begin{center}
\begin{tabular}{|c|c|}
\hline
\rule{0pt}{12pt}\bf{Noise Level}, $\sigma$ & \bf{Error}, $e_{avg}$ \\
\hline
\rule{0pt}{12pt}1 & 66.89 \%\\
\hline
\rule{0pt}{12pt}0.63 & 37.5 \% \\
\hline
\rule{0pt}{12pt}0.46 & 13.14 \% \\
\hline
\rule{0pt}{12pt}0.37 & 0.42 \% \\
\hline
\rule{0pt}{12pt}0.3 & 0.16 \% \\
\hline
\rule{0pt}{12pt}0.26 & 0.16 \%\\
\hline
\rule{0pt}{12pt}0.25 & 0.07 \% \\
\hline
\rule{0pt}{12pt}0.22 & 0.03 \%\\
\hline
\rule{0pt}{12pt}0.2 & 0.01 \% \\
\hline
\end{tabular}
\end{center}
\label{Err}
\end{table}

%
%

\section{Application of the Proposed Approach to Gazepoint Data}
\label{sec:results_real}

  \begin{figure*}
  \centering
  \includegraphics[width=6.5in]{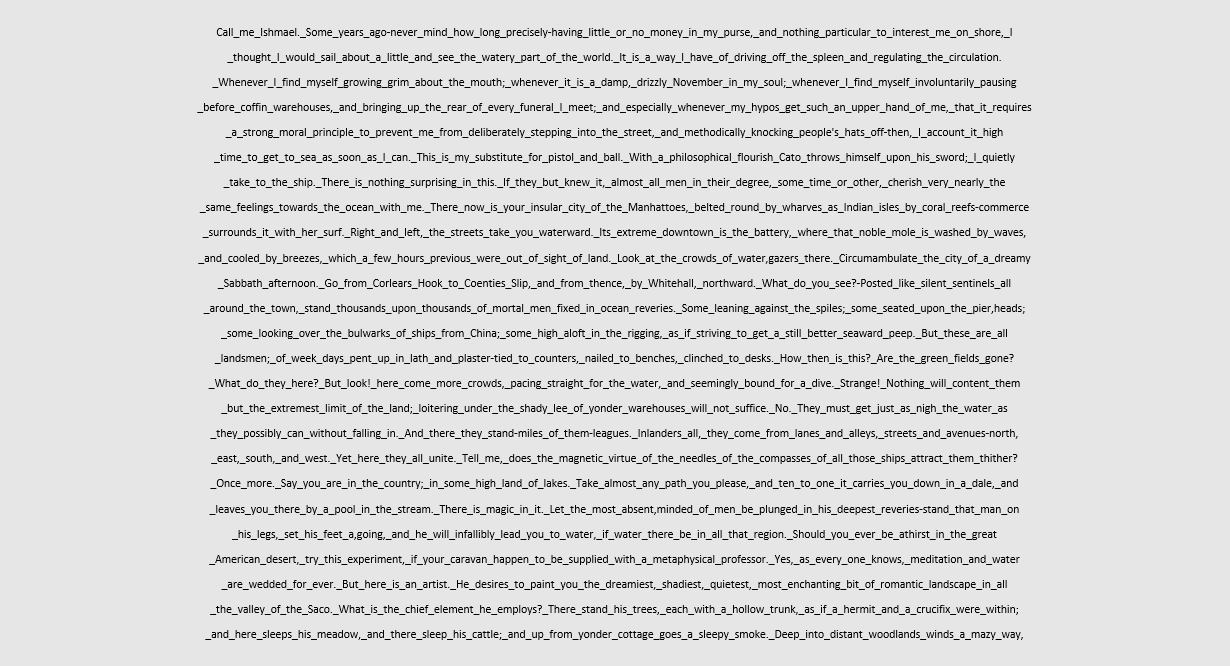}
  \caption{
  {\bf Eye-tracking data collection. }
  A screenshot of the test program, displaying the full 25 lines of text. 
  A human subject will read the lines from top-to bottom while the 
  the eye-tracker collects the gaze fixations of the subject at approximately 60 samples/second. A single line of text at a time is displayed during data collection in order to accurately record the ground truth.}
  \label{fig:gui}
  \end{figure*}

  \begin{figure}
  \centering
  \includegraphics[width=.4\textwidth]{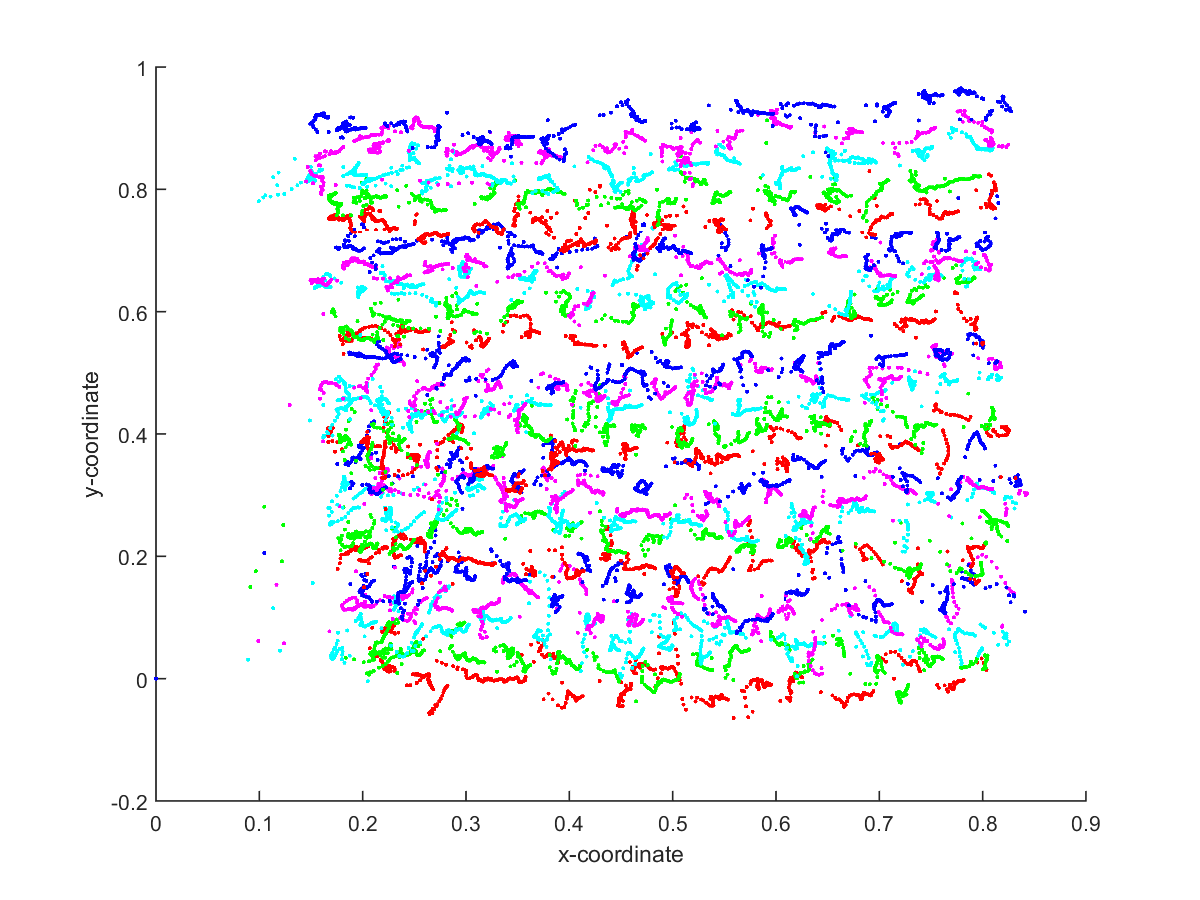}
  \caption{
  {\bf Real eye-tracking data sample.}
  Eye-gaze fixation points for a single page of data, as recorded by the Gazepoint GP3 eye tracker, while the subject is reading a full-page paragraph consisting of 25 lines.}
  \label{fig:realdat}
  \end{figure}
  
 \begin{figure}
  \centering
  \includegraphics[width=.4\textwidth]{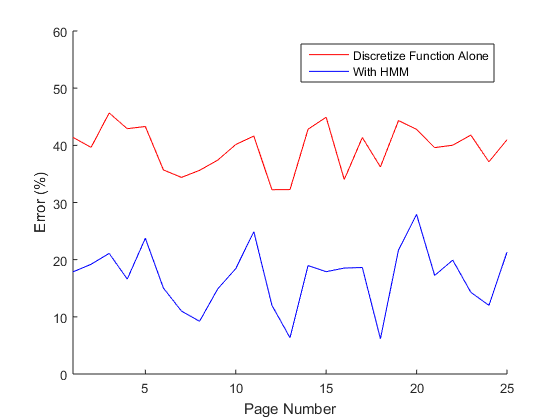}
  \caption{
  {\bf Performance improvement achieved through the proposed method.} Note that in each case, a significant performance improvement is witnessed with the inclusion of the LDS, and detection errors of less than ten percent are achievable. 
  }
  \label{fig:realdat_accuracy}
  \end{figure}

Eye gaze data were collected from a single test subject (a male in his twenties), using a Gazepoint GP3 \cite{gazept} desktop device. A simple data collection program was written in Python, such that communication to and from the Gazepoint device was enabled. The program first initiated a calibration step, which is required by the Gazepoint device before tracking is possible. Upon completion of the calibration step, the test subject was required to press the space key which would simultaneously cue the device to begin logging the $x$ and $y$ eye-gaze fixation coordinates, at 60 Hz, and reveal a single line of text against a solid background, near the top of the display (in this case, a 1920$\times$1080 computer monitor) for which the device was calibrated. Figure \ref{fig:gui} contains a visualization of the display, with each line of text shown, however only a single line of text at a time was displayed during data collection in order to accurately record the ground truth. While the topmost line of text  --- Line 1 --- was displayed, each gaze point corresponding to this line were labeled with a ``1'' to allow for comparison between ground truths and predictions.
 
\par Upon completion of Line 1, the space key was pressed which would cause Line 1 to disappear from view and prompt Line 2 to be displayed on screen at the corresponding location of Line 2, as well as increment the ground truth label to match the current line displayed on screen. In such a manner, eye gaze fixation points were collected for 25 lines of text, which would represent one page worth of data. Figure \ref{fig:realdat} illustrates a typical set of eye gaze fixation points collected for one page of data. Text was fed to the program by a file containing a passage taken from a publicly available copy of Moby Dick by Herman Melville \cite{moby}. 

\par Line predictions were obtained, and the accuracy of the HMM was quantified, by using pocedures consistent to those outlined in previous sections. We collected 25 pages worth of real data at 25 lines per page in order to test the line detection algorithm in a realistic reading setting. To establish a point of comparison, the error between the true states and those predicted by the discretize function alone were determined for each page, and averaged. Similarly, the error between the true states and those predicted by the HMM were determined for each page and also averaged. As is shown in Table \ref{table:realerr}, the addition of an HMM improves prediction accuracy by about 22.5\%, with an average error of about 16.9\%. A comparison of errors for each page is illustrated in Figure \ref{fig:realdat_accuracy}.

\par
\begin{figure}
  \centering
  \includegraphics[width=.4\textwidth]{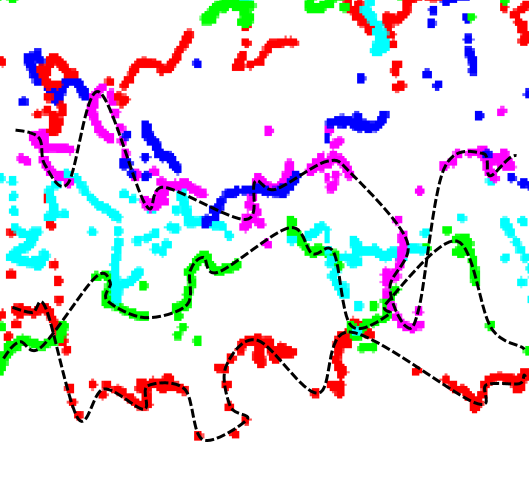}
  \caption{
  {\bf Correlated noise.} 
  A closer look at a portion of the eye-gaze fixation measurements obtained from the Gazepoint eye-tracking device \cite{gazept}.}
  \label{fig:CoNoise}
  \end{figure}

 The noisy nature of the eye-gaze fixation measurements obtained from the eye-tracker was found to be slightly different from the simulated data which was hypothesized in Section \ref{sec:CA} to closely mimic eye-gaze fixation patterns while reading. Referring to Figure \ref{fig:CoNoise}, it can be seen that there is a time correlation in the measurement noise in both the $x$ and $y$ directions in the case of the real data. It must be noted that the difference between two adjacent measurements is approzimately $\frac{1}{60}$ seconds and that the objective is only to track reading progression. Proper HMM modeling for this type of data will require the consideration of sampling time adjustments \cite{rabiner1990tutorial}. 

\begin{remark}[HMM Parameters for Real Data]
In order to keep the HMM simplified, it was decided that the HMM would be trained using training data generated according to the simulated data, described in Section \ref{sec:CA}, at the appropriate level of noise for the particular eye tracking device. For real data, gaze fixation point observations could potentially come from anywhere on the page at any time, and parameters were adjusted to account for this while maintaining a matrix stochastic. It must be noted that an advanced HMM, one which takes sampling time into consideration of the model, will result in improved line detection accuracy than that which was presented in this paper. 
\end{remark}

  \begin{table}
  \caption{Average Errors}
  \label{table:realerr}
  \begin{center}
  \begin{tabular}{|c|c|}
  \hline
  \rule{0pt}{12pt} \bf{Prediction Method} & \bf{Average Error} \\
  \hline
  \rule{0pt}{12pt}Discretize Function & 39.5\% \\
  \hline
  \rule{0pt}{12pt}HMM & 16.9\% \\
  \hline
  \end{tabular}
  \end{center}
  \end{table}

\section{Conclusions and Discussions}
\label{sec:conclusions}

In this paper, we proposed an approach to track the line upon which a reader's eye-gaze is fixated while reading. 
The proposed approach utilizes commercial eye trackers to measure eye-gaze fixations and employs hidden Markov models for line detection. the proposed Line Detection System (LDS) is demonstrated using commercial eye-tracking devices, and the accuracy of the proposed algorithm is shown to be 83.1\%.

%
%
%
%

\section*{Acknowledgements}
\label{sec:acknowledge}
The authors would like to thank Rajankumar Patel (3rd year Electrical Engineering Student at the University of Windsor) for his assistance in collecting the data from the eye-tracker. 
Dr. Balasingam would like to acknowledge Natural Sciences
and Engineering Research Council of Canada (NSERC) for
financial support under the Discovery Grants (DG) program. 

\bibliography{HMM4gazeTracking}

\end{document}